\documentclass[11pt,twoside]{article}
\usepackage{asp2010}

\resetcounters

\begin{document}

\title{Abundance trends in the inner and outer Galactic disk}
\author{T. Bensby,$^1$ A. Alves-Brito,$^{2,4}$ M.S. Oey,$^3$ D. Yong,$^4$ and J. Mel\'endez$^5$
\affil{$^1$Lund Observatory, Box 43, SE-221\,00 Lund, Sweden\\
$^2$Dpto de Astronom\'ia y Astrof\'isica, Pontificia Univ.
Cat\'olica, Santiago, Chile\\
$^3$Department of Astronomy, University of Michigan, Ann Arbor,
MI, USA\\
$^4$Research School of Astronomy and Astrophysics, ANU,
Weston, Australia\\
$^5$Dpto de Astronomia do IAG/USP, Univ. de S\~ao Paulo, S\~ao Paulo, Brasil}}

\begin{abstract}
Based on high-resolution spectra obtained with the MIKE spectrograph
on the Magellan telescopes we present detailed elemental abundances
for 64 red giant stars in the inner and outer Galactic disk.
For the inner disk sample
(4-7\,kpc from the Galactic centre) we find that stars
with both thin and thick disk abundance patterns are present
while for Galactocentric distances beyond 10\,kpc, 
we only find chemical patterns associated with the local thin disk, even for 
stars far above the Galactic plane.  Our results show that the relative densities 
of the thick and thin disks are dramatically different from the solar 
neighbourhood, and
we therefore suggest that the radial scale length of the thick disk is
much shorter than that of the thin disk. A thick disk scale-length of 
$L_{thick}=2.0$\,kpc, and $L_{thin}=3.8$\,kpc for the thin disk, better
match the data.
\end{abstract}

\section{Introduction}

The inner and outer Galactic disks are the most poorly studied regions
of the Milky Way. Especially the abundance structure of 
the inner disk is largely unknown due to the high interstellar 
extinction and contamination by background bulge stars. There are only a 
few studies of bright hot OB stars \citep[e.g.,][]{daflon2004} and Cepheids
\citep[e.g.,][]{luck2006}. The outer disk is somewhat better
studied using red giants in open clusters
\citep[e.g.,][and references therein]{yong2005,carraro2007,jacobson2011}.
Also OB stars  \citep[e.g.,][]{daflon2004,daflon2004outer}, and Cepheids 
\citep[e.g.,][]{andrievsky2004,yong2006} have been observed in the outer disk,
and \cite{carney2005} observed three outer disk field red giants.
Open clusters, OB stars, and Cepheids are all
tracers of the young stellar population of the disk, and it is therefore unclear wether
the inner and outer disk shows a similar abundance structure as seen in the solar
neighbourhood, where it has been shown that the thin and thick disks
have different abundance trends, metallicity distributions, and 
abundance distributions
\citep[e.g.,][]{fuhrmann1998,bensby2003,reddy2003,bensby2005,bensby2007letter2}.

\begin{figure*}
\centering
\resizebox{0.8\hsize}{!}{
\includegraphics[bb=18 155 592 530,clip]{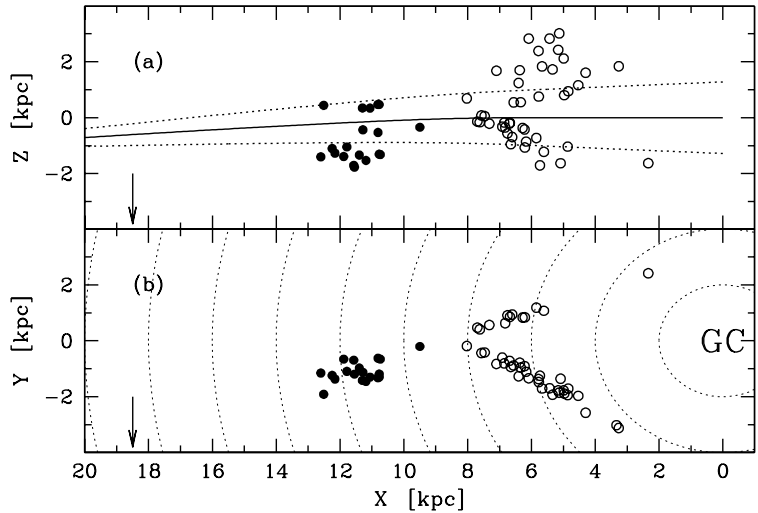}}
\caption{Galactic $X$, $Y$, and $Z$ coordinates.
Dotted lines in (a) represent the
distances above and below the plane where the thin and thick disk 
stellar densities are equal, given the scale-lengths, scale-heights, 
and normalisations for the thin and thick disks given by \cite{juric2008}.
The warp of the disk as given by \cite{momany2006} has been included. 
\label{fig:glonglat} 
}
\end{figure*}

To investigate the abundance structure of these poorly studied regions of the Galaxy,
we have obtained high-resolution and high signal-to-noise spectra of 
44 red giants in the inner disk and 20 red giants in the outer disk 
with the MIKE spectrograph at the Magellan II telescope on Las Campanas in Chile.
The inner disk giants are located at Galactoccentric distances 4-7\,kpc and the outer
disk giants at Galactocentric distances 9-12\,kpc. In order to trace both the thin
and the thick disks, if they are present, the stars were observed at different Galactic
longitudes. Figure~\ref{fig:glonglat} shows the positions of the stars in Galactic
$X$, $Y$, and $Z$ coordinates. The inner disk sample will be valuable 
for verification of the claimed similarities between the nearby thick disk
and the Galactic bulge \citep{bensby2010,alvesbrito2010,bensby2011,gonzalez2011}.
Our results so far have been presented in two letters 
\citep{bensby2010letter,bensby2011letter} and we will in the proceeding give a 
short summary.

\section{Analysis}

Details of the analysis are given in \citep{bensby2010letter}. We would 
like to stress that the analysis is identical to the K giant samples of the 
thin and thick disks in the solar neighbourhood and the bulge K giant
sample presented by \cite{alvesbrito2010}.  Hence, the analysis is strictly
differential between the different stellar populations.

\section{Results and discussion}

\begin{figure*}
\centering
\resizebox{0.9\hsize}{!}{
\includegraphics[bb= 18 170 592 375,clip]{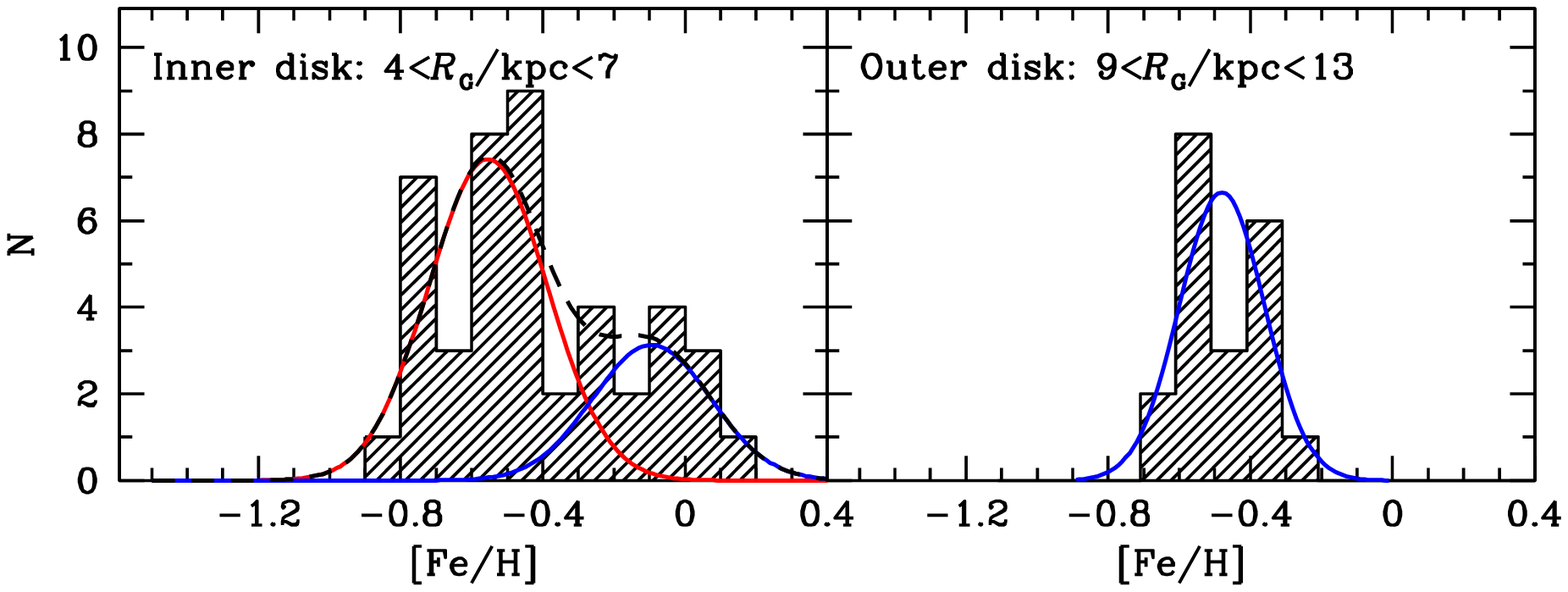}}
\resizebox{\hsize}{!}{
\includegraphics[bb= 0 160 445 620,clip]{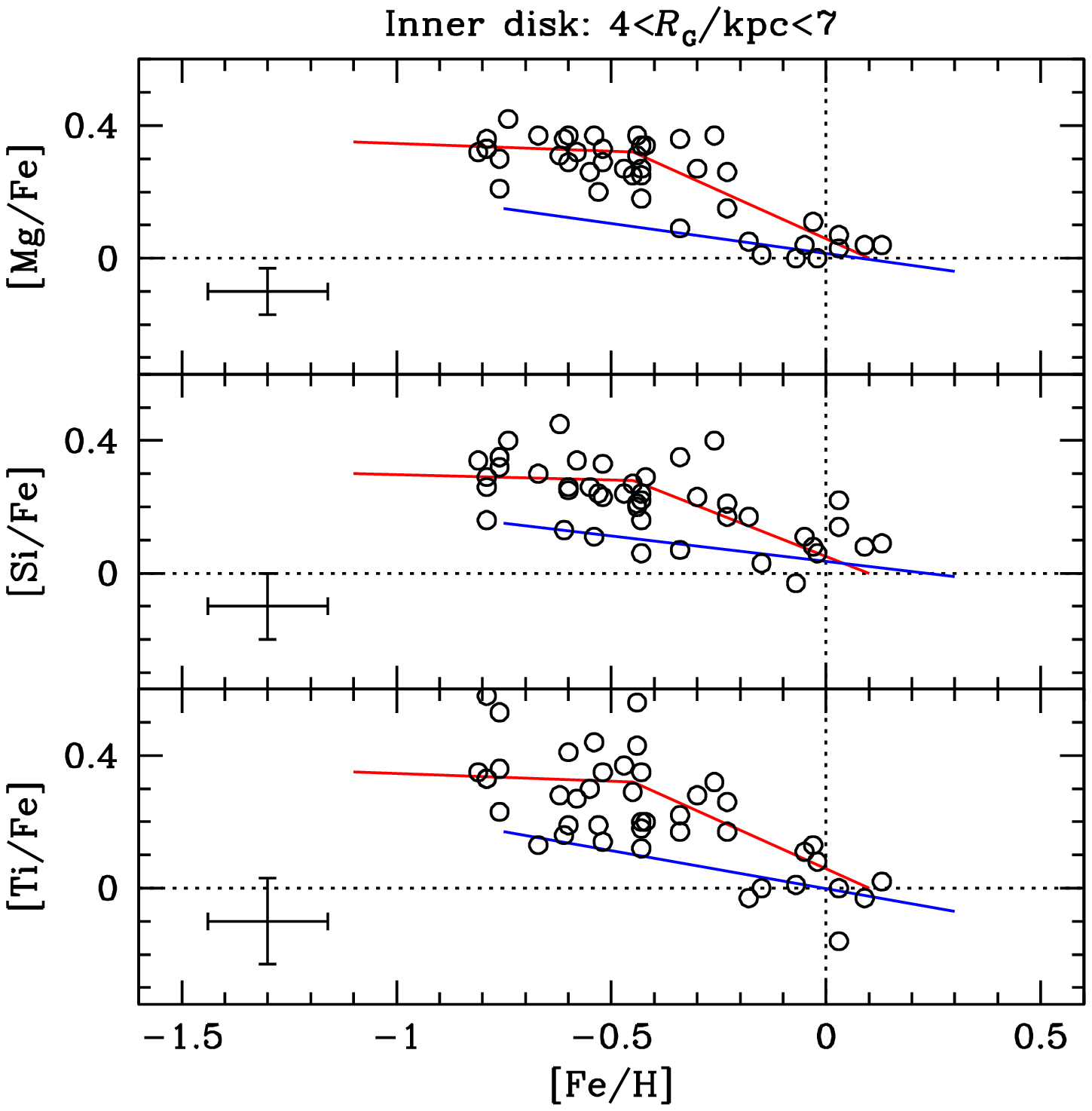}
\includegraphics[bb=75 160 445 620,clip]{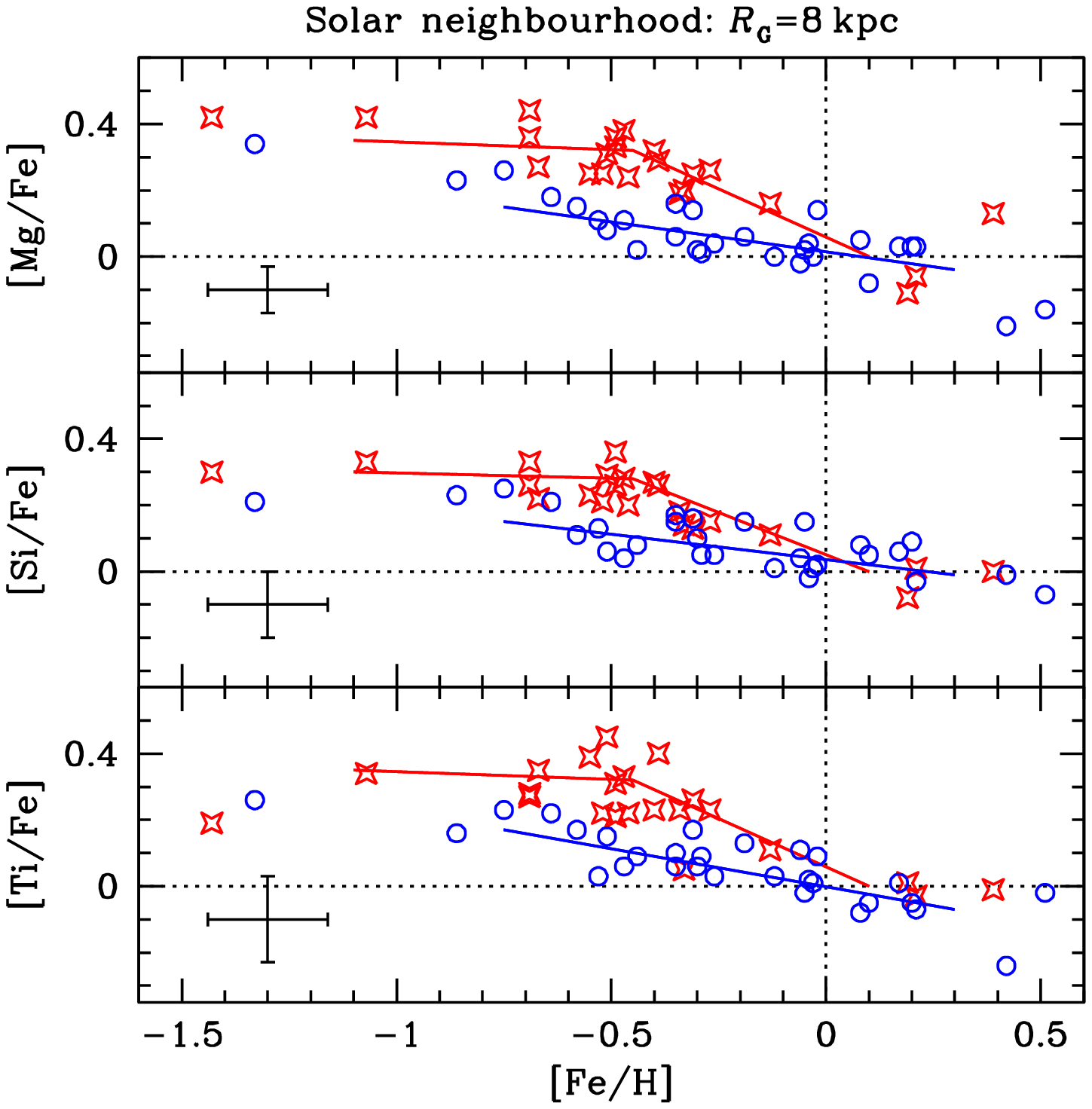}
\includegraphics[bb=75 160 465 620,clip]{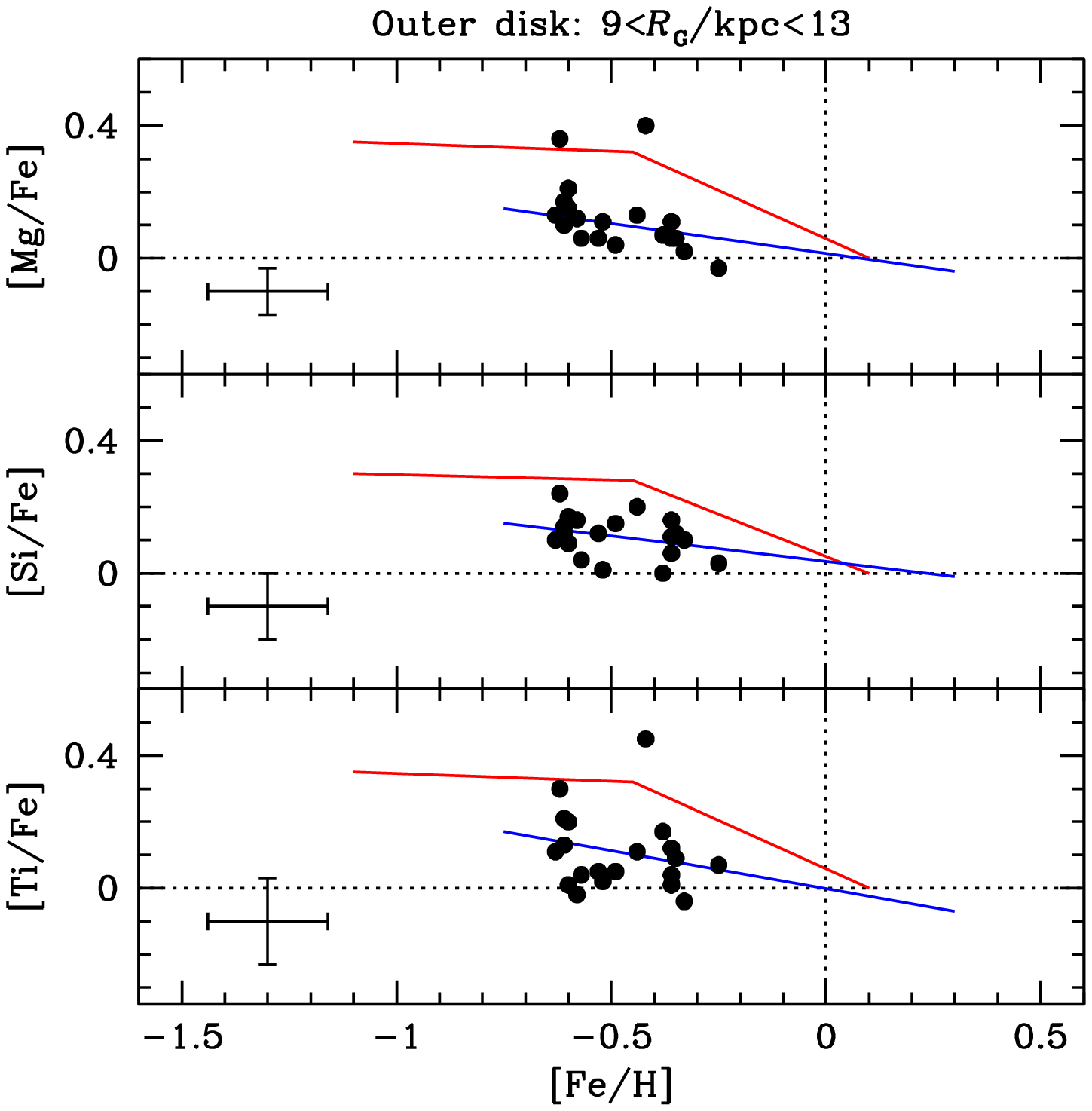}}
\caption{
Abundance trends for the $\alpha$-elements Mg, Si, and Ti. 
The left panel shows the 44 inner disk red giants, the middle 
panel the solar neighbourhood
thin and thick disk stars 
(blue circles and red stars, respectively) by \cite{alvesbrito2010},
and the right panel the 20
outer disk red giants.  Red and blue lines are fiducial lines based 
on the solar neighbourhood samples. 
\label{fig:haltplottar} 
}
\end{figure*}

In the upper
panel of Fig.~\ref{fig:haltplottar} we show the metallicity distributions 
of the inner and outer disk samples. 
A first thing to notice is that the metallicity distributions (MDF)
for the inner and outer disk samples are very different. The inner
disk MDF has a large spread ($\rm \langle[Fe/H]\rangle_{inner}=-0.42\pm 0.27$)
and suggests a bi-modal distribution, while the outer disk
MDF has a much smaller spread 
$\rm \langle[Fe/H]\rangle_{outer}=-0.48\pm 0.12$. 
Within the limited sample, the outer disk MDF is entirely consistent with 
a single value! The dispersion can be attributed solely to 
measurement uncertainties. 
Dividing the inner disk sample into two, one with stars that have
$\rm [Mg/Fe]\geq0.2$ (thick disk) and one with stars that have $\rm [Mg/Fe]<0.2$
(thin disk), results in two metallicity distributions with 
$\rm \langle[Fe/H]\rangle_{inner}=-0.55\pm 0.12$
and $\rm \langle[Fe/H]\rangle_{inner}=-0.09\pm 0.17$, respectively.

\begin{figure}
\centering
\resizebox{0.8\hsize}{!}{
\includegraphics[bb=18 155 592 490,clip]{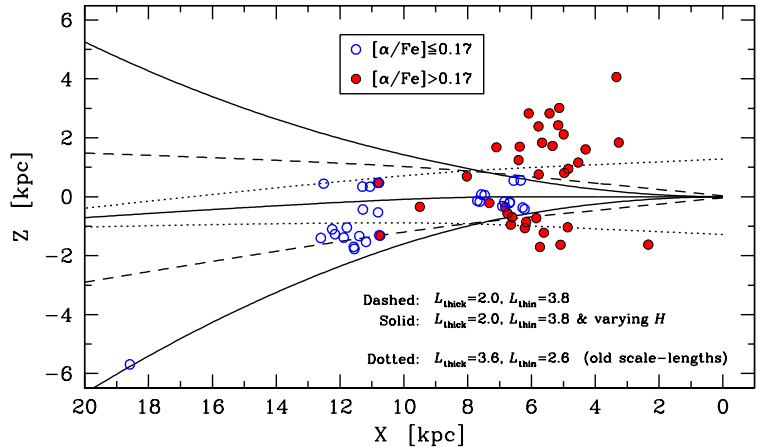}}
\caption{
The inner and outer disk samples in the Galactic $X-Z$
coordinate system. The curves show loci where the two populations
have equal stellar density, for different assumed scale lengths $L$ and
scale heights $H$ Stars
with $\rm [\alpha/Fe]>0.17$ are marked by filled red circles, while stars
with  $\rm [\alpha/Fe]\leq0.17$ by open blue circles.
\label{fig:scalelength} 
}
\end{figure}

The abundance results for the inner and outer disk giants are shown 
in the lower panels of Fig.~\ref{fig:haltplottar}, where they are
compared to the \cite{alvesbrito2010} sample of
thin and thick disk red giants in the solar neighbourhood. 
Regarding the outer disk, almost all stars have abundance ratios 
similar to what is seen in the nearby thin disk. This result is surprising 
because, based on the kinematics and the distances from the plane, 
a majority of the 20 stars should be thick disk 
stars. But only one, or maybe two, of the outer disk giants show thick 
disk abundance patterns. 
The abundance trends of the inner disk sample appears to contain 
stars with abundance patterns consistent with the nearby thin and thick disks. 
An explanation for the apparent lack of stars with thick disk properties
in the outer disk is that the thick disk has a significantly shorter 
scale-length than the thin disk. This would lead to the conclusion that the
thick disk will be more dominant in the inner disk region, and the thin disk
will be more dominant in the outer disk region. 
This is illustrated by the dashed line in 
Fig.~\ref{fig:scalelength} where the thick disk scale-length has been 
changed so that a majority of the outer disk stars are within the
limits where the thick disk stars start to dominate. The solid lines represent 
the case when the scale-height is allowed to vary with Galactocentric radius
(see \citealt{bensby2011letter}).
With the new scale-lengths there is a better match of chemistry vs. vertical
distance from the Galactic plane, for both inner and outer disk samples.

Further discussions and details can be found in \cite{bensby2010letter,bensby2011letter}.

\acknowledgements T.B. was funded by grant No. 621-2009-3911 from The 
Swedish Research Council. This work was also funded by the NSF
grant AST-0448900 to M.S.O. A.A.-B. acknowledges grants from FONDECYT 
(process 3100013). 
\bibliographystyle{asp2010}
\bibliography{referenser}

\end{document}